\documentclass[superscriptaddress,aps,prl,twocolumn]{revtex4-1}
\usepackage{graphicx}
\usepackage{textcomp} \usepackage{amssymb} \usepackage{amsmath}
\usepackage[utf8]{inputenc}
\renewcommand\vec[1]{\mathbf{#1}}
\begin{document}
\title{Flagella-like beating of a single microtubule}
\author{Andrej Vilfan}\thanks{These authors contributed
  equally} \affiliation{Max Planck Institute for Dynamics and
  Self-Organization (MPIDS), 37077 G\"ottingen,
  Germany}
\affiliation{Jo\v{z}ef Stefan Institute, 1000 Ljubljana, Slovenia}
\author{Smrithika Subramani}\thanks{These authors contributed
  equally} \affiliation{Max Planck Institute for Dynamics and
  Self-Organization (MPIDS), 37077 G\"ottingen,
  Germany}
\author{Eberhard Bodenschatz} \affiliation{Max Planck Institute for
  Dynamics and Self-Organization (MPIDS), 37077 G\"ottingen,
  Germany}
\affiliation{Institute for Dynamics of Complex Systems,
  Georg-August-University G{\"o}ttingen, 37073 G{\"o}ttingen, Germany}
\affiliation{ Laboratory of Atomic and Solid-State Physics,
  Cornell University, Ithaca, NY 14853, United States} \author{Ramin
  Golestanian} \affiliation{Max Planck Institute for Dynamics and
  Self-Organization (MPIDS), 37077 G\"ottingen,
  Germany}
\affiliation{Rudolf Peierls Centre for Theoretical Physics,
  University of Oxford, Oxford OX1 3PU, United Kingdom}
\author{Isabella Guido} \email{isabella.guido@ds.mpg.de}  \affiliation{Max Planck
  Institute for Dynamics and Self-Organization (MPIDS), 37077
  G\"ottingen, Germany}
\begin{abstract}
  Kinesin motors can induce a buckling instability in a microtubule
  with a fixed minus end. Here we show that by modifying the surface
  with a protein-repellent functionalization and using clusters of
  kinesin motors, the microtubule can exhibit persistent oscillatory
  motion, resembling the beating of sperm flagella. The observed
  period is of the order of 1 min. From the experimental images we
  theoretically determine a distribution of motor forces that explains
  the observed shapes using a maximum likelihood approach. A good
  agreement is achieved with a small number of motor clusters acting
  simultaneously on a microtubule. The tangential forces exerted by a
  cluster are mostly in the range $0-8\,\rm pN$ towards the
  microtubule minus end, indicating the action of 1 or 2 kinesin
  motors. The lateral forces are distributed symmetrically and mainly
  below $10\,\rm pN$, while the lateral velocity has a strong peak
  around zero. Unlike well-known models for flapping filaments,
  kinesins are found to have a strong ``pinning'' effect on the
  beating filaments. Our results suggest new strategies to utilize
  molecular motors in dynamic roles that depend sensitively on the
  stress built-up in the system.
\end{abstract}
\maketitle
\section{Introduction}
The propulsion of motile cells such as sperms rely on undulating
bending of flagella, appendages of the cellular body able to perform
periodic oscillations. This beating is driven by the interaction
between microtubules and motor proteins and the mechanism that
regulate these interactions, as well as, how they can orchestrate
complex vital processes are not well understood yet.
The functional reactions of living systems operating out of
equilibrium are the result of component rearrangements due to the
intrinsic stochasticity in the system itself. Building a synthetic
molecular system in which components continuously rearrange and
reorganize in the presence of energy sources would enable a deep
understanding of these molecular interactions.

Here we present experimental and theoretical results on a minimal
system made of a single microtubule with a fixed end and a small
number of kinesin-1 motor proteins (kinesin hereafter) that in the
presence of ATP perform a continuous motion resembling the beating of
sperm flagella.

In the past decades these components were used in \textit{in vitro}
gliding assay \cite{Kron,Vale} where the motor proteins were attached
on the surface of a glass coverslip at high density. Directed sliding
motion of filaments was enabled when they came into contact with the
surface bound motors.
Previous studies showed microtubule buckling in a low density gliding
assay due to a single kinesin motor \cite{howard96}, waving and
rotating motion of biofilaments in gliding assay
\cite{Bourdieu,Sekimoto.Toyoshima1995} as well as spiral formation of
microtubules caused by pinning at the leading end \cite{Bourdieu} and
by the interactions with neighboring microtubules
\cite{Ross}. 

In this study we quantitatively analyze the buckling instabilities of
single microtubules clamped at one end and subjected to the forces
exerted by motor proteins. We found that by using a specific surface
treatment for the substrate and organizing the motors in randomly
distributed clusters we obtain filaments that perform continuous
beating.
	
\begin{figure}
  \includegraphics[width=\linewidth]{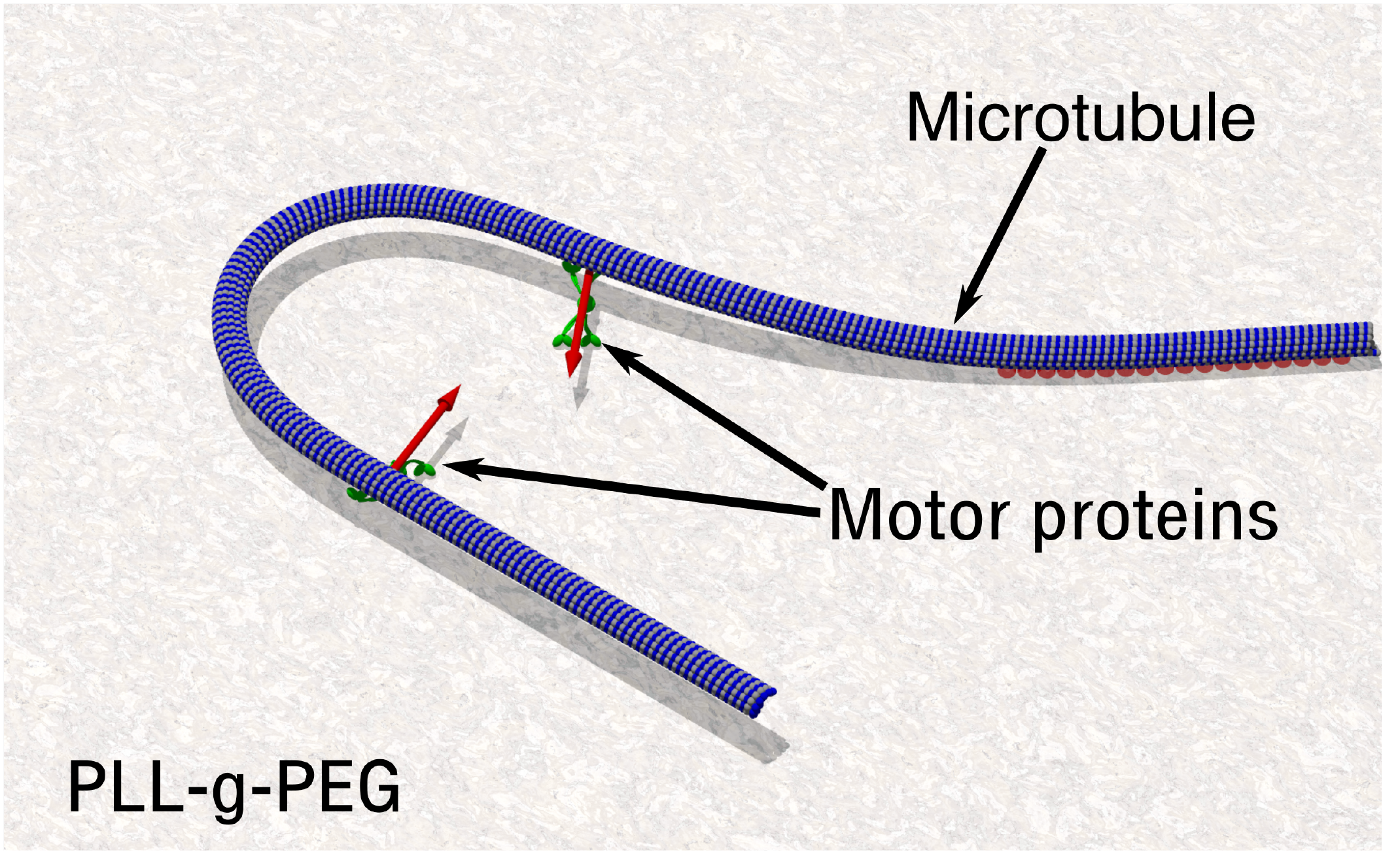}
  \caption{Schematic representation of the minimal system. A single
    filament clamped at one end beating and buckling under the action
    of motor proteins interacting with PLL-g-PEG functionalized
    surface.}
  \label{fig:3D}
\end{figure}

We can describe the system as follows (Fig.~\ref{fig:3D}): (i) the
kinesin motors in the cluster simultaneously bind either to the
microtubules or unspecifically to the surface; (ii) the motors exert
forces on the microtubule which are directed along the tangential
direction. Above a threshold the straight configuration of the
filament becomes unstable inducing the buckling of the filament; (iii)
the force that bends the filament has a longitudinal component as well
as a perpendicular component that allows the microtubule to snap back
to its straight configuration. This can be explained by either the
filament detaching from the motors or by the motors breaking the bond
with the protein-repellent surface and being pulled away while
remaining attached to the filament. Both lead to a subsequent buckling
when the filament reaches the new position.
The behaviour of the system is self-organized by the elasticity of the
filament, the active forces exerted by motor proteins and by using
attachment and detachment between the building blocks of this minimal
system.
	
\section{Results and discussion}
\begin{figure*}
  \begin{center}
    \includegraphics[width=\linewidth]{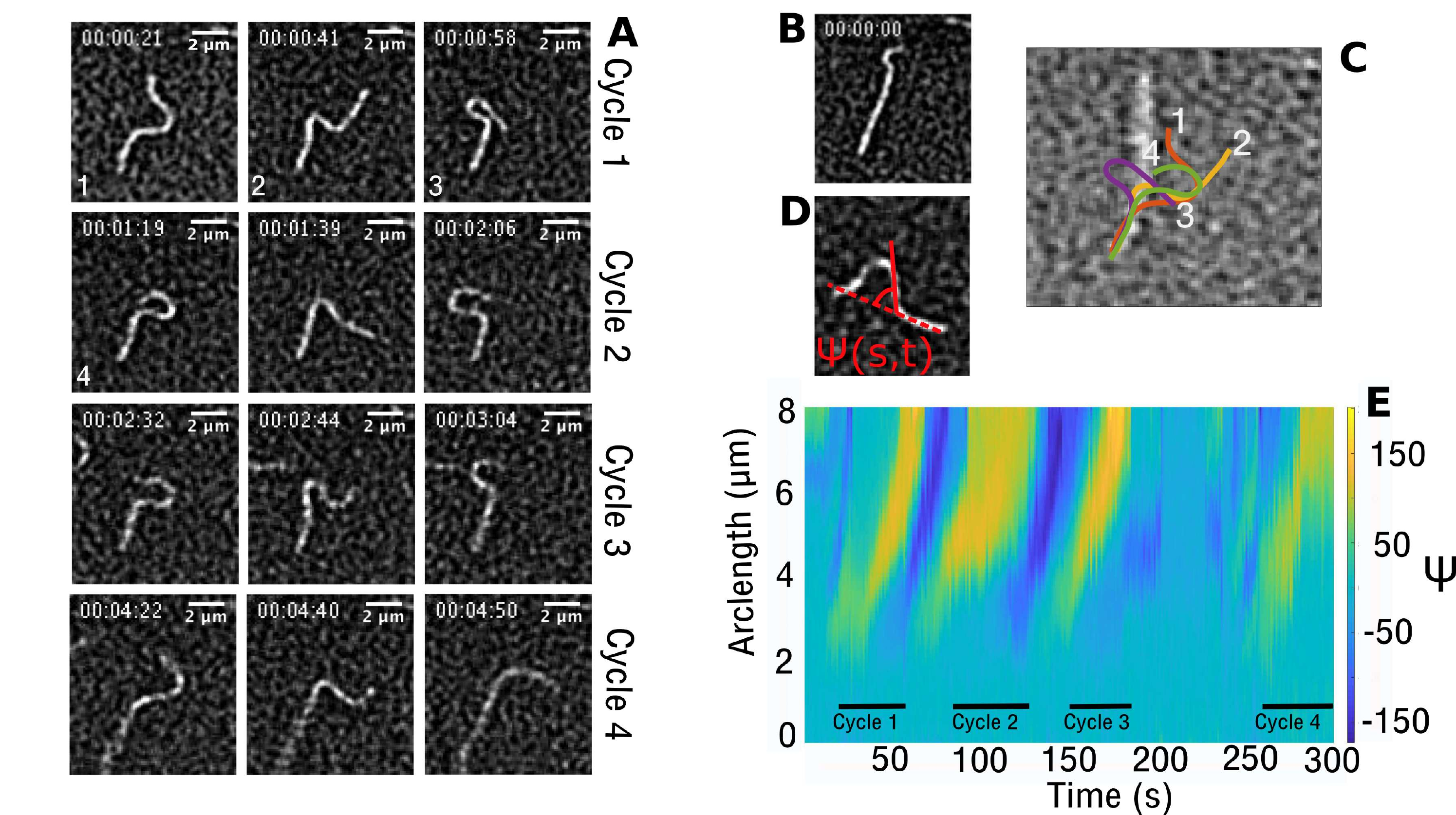}
    \caption{A. Oscillation cycles of a single filament clamped on the
      surface over time starting at t= 0 from the configuration shown
      in B. B. Starting position of clamped filament. C. Filament
      tracking shows the time evolution of the emergent
      oscillations. For visualization reason only the tracking of
      first cycle is shown. The numbers indicate the chronology of the
      oscillation as well as correspond to the filament positions
      shown in A (see numbers on the bottom left of the
      micrographs). D. Tangent angle $\psi$(s,t) as a function of
      arclength s along the flagellum. D. Kymograph of the tangent
      angle over time along the filament. The oscillation and its
      duration corresponding to each cycle is pointed out in the
      figure.}
    \label{fig:tangent}
  \end{center}
\end{figure*}

\paragraph{Experimental results}
Here, we report the quantitative analysis of buckling instabilities of
clamped microtubules caused by compressive forces exerted by kinesin
motors.  Diluted polymerized microtubules of approx. 8 $\mu$m length
were decorated with motor protein clusters by mixing them with
biotinylated kinesin clustered by using
streptavidin. 
The surface was functionalized by using PLL-g-PEG (see Materials and
Methods section). This nonionic polymer has been proven to limit
biological interactions \cite{Spencer} and in this set-up it prevents
irreversible protein adsorption on the surface. A sample of the
microtubules-kinesins mixture was poured on the PLL-g-PEG modified
cover slip and the experimental chamber was sealed in order to avoid
any artefacts caused by fluid streaming.
In our experiments the chosen functionalization of the glass surface
almost entirely avoids the microtubules adsorption. However, we could
occasionally observe microtubules with one tip clamped (typically 2
$\mu$m) to the surface (1-2 samples in each experiment) likely due to
electrostatic interaction with the polylysine backbone of the
PLL-g-PEG functionalization. The anchor point had no rotational or
translational degree of freedom. The free part of the microtubule was
observed to perform oscillatory motion as shown in
Fig.~\ref{fig:tangent}. This motion is attributed to the kinesin
motors which could bind with their heads attached to the microtubule
and unspecifically bound to the surface simultaneously, due to their
organization in clusters. When the kinesin-clusters were bound in this
manner the motors transiently pushed the filament forward resulting in
cyclic conformational changes for time intervals up to 5 minutes.
The oscillations of the microtubule can be divided into cycles,
defined as the time that the filament needs to move from one portion
of the space to the adjacent one and to come back
(\ref{fig:tangent}-A). The filament tracking of one cycle is exemplary
shown in figure \ref{fig:tangent}-C where the four subsequent
positions are overlaid and labeled in order to show the beating
behaviour of the microtubule.
	
The ability to beat and buckle persistently depends on (i) an optimal
concentration of motor clusters and (ii) an interacting surface with a
drastically reduced protein adsorption. Control experiments carried
out without a PLL-g-PEG surface functionalization showed no buckling
filaments, proving that the unspecific binding of both motor proteins
and microtubules does not allow any filament to move freely.

The motor concentration needed to buckle the filament can be estimated
from the following simple argument. A filament of length $L$ is
attached to motor clusters with an average spacing $\lambda$, which we
treat as pinning points. If we assume that the filament is torque-free
at the motor positions (which is a simplification), the force needed
to buckle the segment is
\begin{equation}
  \label{eq:buckling}
  F_B=\frac{\pi^2 EI}{\lambda^2}\;.
\end{equation}
The compressive force, exerted by the ensemble of motors, is highest
in the segment next to the clamped end, and corresponds to
$F_M L/\lambda$, where $F_M$ is the stall force of a motor
cluster. Both become equal at a critical length
\begin{equation}
  \label{eq:lcrit}
  L_C=\frac{\pi^2 EI}{F_M \lambda}\;,
\end{equation}
which is inversely proportional to the motor spacing or proportional
to their density. With the values $F_M=8\,\rm pN$,
$EI=0.4\times 10^{-23}\,\rm Nm^2$ and $\lambda=1\,\rm \mu m$, we
obtain $L_C=5\,\rm \mu m$. A filament of this length will generally
buckle if the number of motors acting on it is in the range 1-5.
	
We verified the importance of the motor protein concentration by
repeating the experiments with a different motors concentration. The
experiments showed a reduced capacity to buckle continuously.  By
using a 100-fold higher concentration not continuous oscillations were
observed.
The buckling events were less frequent and limited in time compared to
the analysed case (movie S2). By reducing 2-fold the concentration
again the buckling events were less frequent and limited in time as
the sliding of filaments was followed by the filaments leaving the
focal plane due to the missing grasp (movie S3).

We analyze the movement pattern of our filaments. Their beating was
observed to be quasi 2D and this allowed us to track the filament
shape. We characterized the tracked filament shape by a tangent angle
$\psi$(\textit{s,t}), which describes the angle between the straight
line representing the initial position of the filament and the local
tangent of the arclength \textit{s} along the filament at the time
\textit{t} with $ 0\leqslant \textit{s}\leqslant \textit{L}$ and
\textit{L} the length of the free moving filament portion
(Fig.~\ref{fig:tangent}-D).  We tracked the shape changes for
\textit{n} frames with \textit{n} up to 300.
We obtained a matrix $\psi$(\textit{s,n}) that represents the
kymograph of the filament movement (Fig.~\ref{fig:tangent}-D).  We can
observe a pattern indicating a wavy motion propagating from one end of
the filament along its entire length and reflecting the periodicity of
the filament beating, i.e. the four cycles which its movement can be
divided into, as defined above and depicted in
Fig.~\ref{fig:tangent}-A.  Thus, the beating profile shown in
Fig.~\ref{fig:tangent}-E, is similar to the beating of eukaryotic
flagella.

The analogy between filaments subject to tangential compressive forces
and beating flagella has already been pointed out in several
theoretical studies. The basic problem consists of a semiflexible
polymer with one clamped end, subject to local drag and an active
tangential force. Sekimoto et al. \cite{Sekimoto.Toyoshima1995} showed
the existence of a dynamical instability, above which the filament
undergoes flapping motion, whereas Bourdieu et al. \cite{Bourdieu}
briefly presented a waving motion in actin and microtubules. A related
problem with a single follower force at the free filament tip was
solved in Ref. \cite{DeCanio.Goldstein2017}. If the filament is free to
move in 3 dimensions, there is an additional spinning
state \cite{Ling.Kanso2018}. If the clamped end is replaced by a cargo,
resembling a sperm head, the numerically obtained solutions show
rotation or sperm-like beating \cite{Gompper}. Despite the
similarities, there are also essential differences between our system
and these models. The small number of motors involved makes the
dynamics less deterministic, while the ability of kinesins to sustain
off-axis loads largely reduces the lateral motion of the filament,
except during periods of fast relaxation, presumably caused by the
detachment of motors from the distal region.

We have the conjecture that the continuous buckling dynamics might be
also due to the reversible attachment of the motors to the surface,
which was tailored to allow a binding that is sufficient stable for
activating the 'walk' on the microtubule but so weak to dissociate
under mechanical stimulus. In this way the system self-organized by
using the force of motor proteins that continuously rearranged by
breaking and reforming bonds to the surface. The weak bond to the
surface was tested by applying a fluid flow at shear rate estimated as
$4\times 10^{4}\, \rm s^{-1}$ that was able to wash out the motors
unspecifically bound to the surface.  Other studies show that surface
modifications are able to create reversible motors binding responsible
for self-organization of microtubules dynamics due to continuous
rearrangement of motors distribution \cite{Hess}.  We believe that the
force exerted by the motors in one portion of the filament inducing
its buckling reached a threshold above which it was able to pull away
the other motors from the surface.  As soon as the microtubule was not
under the constrain of the motors totally or partially relaxed to the
straight position due to its elastic properties transporting the
motors in another position. After this rapid position change the
reattachment of the motors occurred again and a subsequent new sliding
of the filament.

\begin{figure*}
  \begin{center}
    \includegraphics[width=0.9\linewidth]{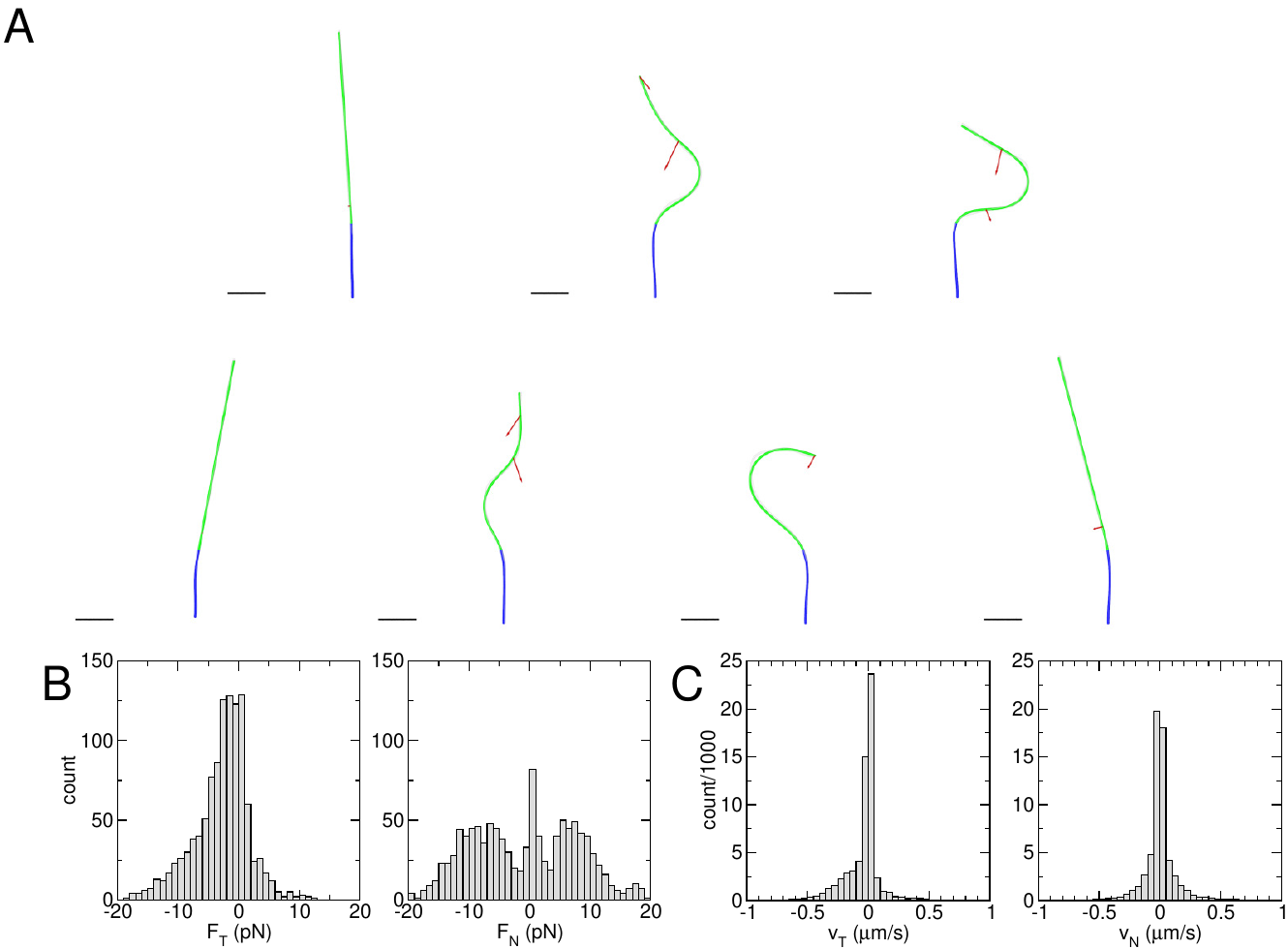}
    \caption{A. Reconstruction of motor forces (red arrows) from
      filament shapes (gray: experimental; green: fitted; blue: fixed
      filament end). Scale bar: $1\,\rm\mu m$, $10\,\rm pN$.
      B. Distribution of tangential (T) and normal (N) force
      components. A negative sign ($F_T<0$) represents a force pushing
      the microtubule towards the clamped end. C. Tangential and
      normal velocity components of points on the filament. $v_T<0$
      indicates the filament moving towards the clamped end. The
      histograms were obtained from the analysis of three continuously
      beating microtubules over time intervals ranging from 1:30 to 5
      minutes.  }
    \label{fig:reconstruction}
  \end{center}
\end{figure*}

We used a maximum likelihood method to reconstruct a distribution of
motor forces that gives the observed filament shape. Examples of
filaments with corresponding forces are shown in
Fig.~\ref{fig:reconstruction}. The tangential components of the forces
(Fig.~\ref{fig:reconstruction}-B) shows a strong accumulation in the
range between $-8\,\rm pN$ and $0\,\rm pN$ (a negative sign denotes
forces pushing the filament towards the clamped end, which is the MT
minus end), consistent with the forces produced by single kinesin
motors. Larger values, up to $-20\,\rm pN$, could result from the
action of multiple motors, either within a cluster, or sufficiently
close together that they become unresolvable. Normal forces are mostly
in the range $-10$ to $10\,\rm pN$, also consistent with 1-2
kinesins. The tangential velocity of the filament shows a highly
skewed distribution, mainly in the range $-400\,\rm nm/s$ to
$0\,\rm nm/s$ (negative sign shows the filament moving towards the
clamped end, consistent with the action of kinesins). However, both
velocities sometimes exceed $2\,\rm \mu m/s$, namely as the free
filament end snaps to a relaxed position.

In conclusion, we report an example of autonomous molecular system
that dynamically self-organizes through its elasticity and the
interaction with the environment represented by the active forces
exerted by motor proteins resembling the beating of a flagellum.
Assembling such minimal systems that can mimic the behaviour of much
more complex biological structure might help to unveil the basic
mechanism underlying the beating of real cilia and flagella.

\section{Materials and Methods}
\textit{\textbf{Non-Adsorbing Surface Coatings and experimental
    chamber assembly}}. Glass coverslips ($64 \times 22\, \rm mm^2$,
VWR) were cleaned by washing with 100\,$\%$ ethanol and rinsing in
deionized water. They were further sonicated in acetone for 30 min and
incubated in ethanol for 10 min at room temperature. This was followed
by incubation in a 2\,$\%$ Hellmanex III solution (Hellma Analytics)
for 2h, extensive washing in deionized water and drying with a
filtered airflow.  The cleaned coverslips are immediately activated in
oxygen plasma (FEMTO, Diener Electronics, Germany) for 30 s at 0.5
mbar and subsequently incubated in 0.1 mg/mL
Poly(L-lysine)-graft-poly(ethylene glycol) (PLL-g-PEG) (SuSoS AG,
Switzerland) in 10 mM HEPES, pH = 7.4, at room temperature for 1 h on
parafilm (Pechiney, U.S.A.).  Finally, the coverslips were lifted off
slowly and the remaining PLL-g-PEG solution was removed for a complete
surface dewetting.  The experimental chamber was obtained by cutting a
window (8 mm$\times$8 mm) on double-sided tape of thickness 10 $\mu$m
(No. 5601, Nitto Denk Corporation, Japan) and sandwiched between two
functionalized coverslips.

\textit{\textbf{Kinesin-1 Decorated Microtubules Experiments.}}
Microtubules were polymerized from 2.7 mg/ml
HiLyte\textsuperscript{TM} labeled porcine brain tubulin
(Cytoskeleton, Inc., U.S.A.) in M2B (80 mM PIPES, adjusted to pH = 6.9
with KOH, 1 mM EGTA, 2 mM MgCl\textsubscript{2}) with 5 mM
MgCl\textsubscript{2}, 1 mM GTP, 5\,$\%$ DMSO at 37 \textdegree C for
30 min. The microtubules were stabilized and diluted 2000-fold in M2B
containing 7 $\mu$M taxol. The plasmid that codes biotin-labeled
kinesin 401 (K401) was a gift from Jeff Gelles (pWC2 - Addgene plasmid
\# 15960 ; http://n2t.net/addgene:15960; RRID\textunderscore
Addgene\textunderscore15960) \cite{Subramanian445}. Kinesin 401 was
purified as previously published \cite{Gilbert, Young} and the
kinesin-streptavidin complexes were prepared by mixing 0.2 mg/ml
kinesin 401, 0.9 mM dithiothreitol (DTT), 0.1 mg/ml streptavidin
(Invitrogen, S-888) dissolved in M2B and incubated on ice for 15 min.
4 $\mu$l of this mixture was mixed with 0.5 mg/ml glucose, 0.65 mM
DTT, 0.2 mg/ml glucose oxidase (Sigma G2133), 0.05 mg/ml catalase
(Sigma C40), 1 mM ATP, 1.7 $\mu$l pyruvate kinase/lactic dehydrogenase
(PK/LDH, Sigma, P-0294), 32 mM phosphoenol pyruvate (PEP, VWR
AAB20358-06), 2.4 mM Trolox (Sigma 238813) to form an active clusters
solution.  The microtubules and active clusters solution in different
concentrations were mixed 15 min before use according to the intended
experiments. 2 $\mu$l of this mixture were pipetted onto the PLL-PEG
functionalized glass and the experimental chamber was sealed.

\textit{\textbf{Imaging and Tracking}}. Image acquisition was
performed using an inverted fluorescence microscope Olympus IX81
(Olympus, Japan) with a 63× oil-immersion objective (Olympus,
Japan). For excitation, a Lumen 200 metal arc lamp (Prior Scientific
Instruments, U.S.A.) was applied. The data was recorded with a
Photometrics Cascade II EMCCD camera (512 $\times$ 512 px). The images
were acquired every 1 s with an exposure time of 100 ms. The
microtubule trajectory was manually tracked by tracing the filament
contour using JFilament, an ImageJ plugin for segmentation and
tracking \cite{Smith.Vavylonis2010}.

\textit{\textbf{Reconstruction of Kinesin Forces.}}  To determine a
minimal configuration of forces that can lead to the observed shapes,
we maximized the quantity
\begin{equation}
  \ell=-\int_{L_0}^L (\Delta(s))^2 ds - \mu \sum_{i=1}^{N_f} {\vec f}_i^2- \nu N_f\;.
\end{equation}
Here $\Delta$ denotes the shortest distance between a point on the
experimental and the shape, calculated from the forces. The integral
runs over the part of the filament that is free to move, as obtained
from an analysis of lateral fluctuations. The second and the third
term were introduced to avoid overfitting: they penalize solutions
with excessive forces or with too many parameters. For a collection of
forces, $\vec f_1 ,\ldots,\vec f_{N_f}$, located at positions
$s_1,\ldots s_{N_f}$, the expected filament shape is calculated from
the equations
\begin{align}\label{eq:m}
  \frac{d\vec M}{ds}&=- {\vec t} \times \left[ \sum_{i=1}^{N_f} \Theta(s_i-s) {\vec f}_i\right] \\
  \frac{d\vec t}{ds}&= \frac 1 {EI} \vec{t} \times \vec{M}\\
  \label{eq:x}          \frac{d\vec x}{ds}&= \vec t
\end{align}
with the boundary condition ${\vec M} (L)=0$. We assume a planar
filament shape, such that $\vec f_i$, $\vec x$ and $\vec t$ lie in the
horizontal plane and $\vec M$ perpendicular to it. We solve the
differential equations (\ref{eq:m}-\ref{eq:x}) numerically and
evaluate the deviation of the solution from the experimental shape
$\Delta(s)$. Examples of reconstructed force distributions for one
experiment are shown in Fig.~\ref{fig:reconstruction}-A.
\begin{acknowledgements}
  The authors thank the MaxSynBio Consortium which is jointly funded
  by the Federal Ministry of Education and Research of Germany and the
  Max Planck Society.
\end{acknowledgements}
\section*{Supplemental movies}
  \begin{itemize}
  \item S1: Time-lapse movie of the reconstruction of force
    distribution acting on the microtubule.
  \item S2: Time-lapse movie of an experimental assay carried out by
    using a 100-fold higher motor protein concentration.
  \item S3: Time-lapse movie of an experimental assay carried out by
    using a 2-fold lower motor protein concentration.
  \end{itemize}

\end{document}